\documentclass{appolb}\usepackage{epsfig}
\title{Low-x contribution to the Bjorken sum rule within unified 
$ln^2x+$LO DGLAP approximation}

\author{Dorota Kotlorz $^1$, Andrzej Kotlorz $^2$
\address{$^1$Department of Physics Ozimska 75, $^2$Department of 
Mathematics Luboszycka 3, Technical University of Opole, 
45-370 Opole, Poland, e-mail $^1$: {\tt dstrozik@po.opole.pl}}}

\begin{document}
\pagestyle{plain}
\eqsec
\maketitle

\begin{abstract}
The small-$x$ contributions to the Bjorken sum rule within unified picture 
$ln^2x+$LO DGLAP for different input parametrisations $g_1^{NS}(x,Q_0^2)$ 
are presented. Theoretical predictions for $\int_{0}^{0.003} g_1^{NS}(x,Q^2=10)
dx$ are compared with the SMC small-$x$ data. Rough estimation of the slope 
$\lambda$, controlling the small-$x$ behaviour of $g_1^{NS}\sim x^{-\lambda}$ 
from the obtained results and SMC data is performed. The crucial role of the 
running coupling $\alpha_s=\alpha_s(Q^2/z)$ at low-$x$ is taken into account.
\end{abstract}


\section{Introduction}

The results of SIDIS (semi inclusive deep inelastic scattering) experiments
with polarised beams and targets enable the extraction of the spin dependent
quark and gluon densities. This powerful tool of studying the internal spin
structure of the nucleon allows verification of sum rules. One of them is
the Bjorken sum rule (BSR) \cite{b1}, which refers to the first moment of
the nonsinglet spin dependent structure function $g_1^{NS}(x,Q^2)$. Because
of $SU_f(2)$ flavour symmetry, BSR is regarded as exact. Thus all of
estimations of polarised parton distributions should be performed under the
assumption that the BSR is valid. Determination of the sum rules requires
knowledge of spin dependent structure functions over the entire region of
$x\in (0;1)$. The experimentally accessible $x$ range for the spin dependent
DIS is however limited ($0.7>x>0.003$ for SMC data \cite{b2}) and therefore
one should extrapolate results to $x=0$ and $x=1$. The extrapolation to
$x\rightarrow 0$, where structure functions grow strongly, is much more
important than the extrapolation to $x\rightarrow 1$, where structure
functions vanish. Assuming that the BSR is valid, one can determinate from
existing experimental data the very small-$x$ contribution ($0.003>x>0$) to
the sum rule. Theoretical analysis of the small-$x$ behaviour of 
$g_1^{NS}(x,Q^2)=g_1^{p}(x,Q^2)-g_1^{n}(x,Q^2)$ together with the broad
$x$-range measurement data allow verification of the shape of the input
parton distributions. In this way one can determinate the free parameters in
these input distributions. Experimental data confirm the theoretical
predictions of the singular small-$x$ behaviour of the polarised structure
functions. It is well known, that the low-$x$ behaviour of both unpolarised
and polarised structure functions is controlled by the double logarithmic
terms $(\alpha_s ln^2x)^n$ \cite{b3},\cite{b4}. For the unpolarised case, this
singular PQCD behaviour is however overridden by the leading Regge
contribution \cite{b5}. Therefore, the double logarithmic approximation is
very important particularly for the spin dependent structure function $g_1$.
The resummation of the $ln^2x$ terms at low $x$ goes beyond the standard LO
and NLO PQCD evolution of the parton densities. The nonsinglet polarised
structure function $g_1^{NS}$, governed by leading $\alpha_s^n ln^{2n}x$ terms, 
is a convenient function both for theoretical analysis (because of its
simplicity) and for the experimental BSR tests. The small-$x$ behaviour of
$g_1^{NS}$ implied by double logarithmic approximation has a form
$x^{-\lambda}$ with $\lambda\approx 0.4$. This or similar small-$x$ 
extrapolation of the spin dependent quark distributions have been assumed in 
recent input parametrisations e.g. in \cite{b6},\cite{b7},\cite{b14}. More 
singular parametrisation of $g_1^{0NS}(x,Q_0^2=4)\sim x^{-0.8}$ at small-$x$, 
based on the QCD (LO and NLO) analysis of the world data on polarised deep 
inelastic scattering, has been presented in \cite{b15}. 
Mentioned above double logarithmic approach is however inaccurate for QCD 
analysis at medium and large values of $x$. Therefore the double logarithmic 
approximation should be completed by LO DGLAP $Q^2$ evolution. In our 
theoretical analysis within $ln^2x+$LO DGLAP approach we estimate $g_1^{NS}$ at 
low-$x$ and hence the small-$x$ contributions 
$\int_{0}^{x_0} g_1^{NS}(x,Q^2) dx$, $\int_{x_1}^{x_2} g_1^{NS}(x,Q^2) dx$ 
($x_0, x_1, x_2 \ll 1$) to the BSR for different input quark parametrisations: 
the Regge nonsingular one and the singular one. We compare our results with 
the suitable experimental SMC data for BSR. In the next section we recall some 
of the recent theoretical developments concerning the small-$x$ behaviour of 
the nonsinglet polarised structure function $g_1^{NS}$. Section 3 is devoted 
to the presentation of the unified $ln^2x+$LO DGLAP approximation. We also
discuss the role of the running coupling $\alpha_s$. Section 4 contains our 
results for the structure function $g_1^{NS}$ at small-$x$ and for 
contributions to the Bjorken sum rule $\Delta I_{BSR}(x_1,x_2,Q^2)=
\int_{x_1}^{x_2} g_1^{NS}(x,Q^2) dx$ ($x_1, x_2 \ll 1$). We present our 
predictions using flat (nonsingular) $\sim (1-x)^3$ and singular 
$\sim x^{-\lambda}$  at small-$x$ parametrisations of the input structure 
function $g_1^{NS}(x,Q_0^2)$ as well. We compare our results with the SMC data 
for the small-$x$ contribution to the BSR. We roughly estimate the slope 
$\lambda$ controlling the small-$x$ behaviour of $g_1^{NS}\sim x^{-\lambda}$ 
from our $g_1^{NS}$ predictions and from the SMC data, basing on the validity 
of the BSR. We compare also results $\Delta I_{BSR}(x_1,x_2,Q^2)$ and 
$g_1^{NS}(x=10^{-6},Q^2=10)$ in different 
approximations: pure LO DGLAP, pure $ln^2x$, $ln^2x+$LO DGLAP and obtained 
for different $\alpha_s$ parametrisations: $\alpha_s=const$, 
$\alpha_s=\alpha_s(Q^2)$, $\alpha_s=\alpha_s(Q^2/z)$. Finally, Section 5
contains a summary of our paper.

\section{Small-$x$ behaviour of the nonsinglet spin dependent structure
function $g_1^{NS}(x,Q^2)$}

The small value of the Bjorken parameter $x$, specifying the longitudinal
momentum fraction of a hadron carried by a parton, corresponds by definition
to the Regge limit ($x\rightarrow 0$). Therefore the small-$x$ behaviour of
structure functions can be described using the Regge pole exchange model 
\cite{b5}. In this model the spin dependent nonsinglet structure function
$g_1^{NS}=g_1^p-g_1^n$ in the low-$x$ region behave as:
\begin{equation}\label{r2.1}
g_1^{NS}(x,Q^2) = \gamma (Q^2) x^{-\alpha_{A_1}(0)}
\end{equation}
where $\alpha_{A_1}(0)$ is the intercept of the $A_1$ Regge pole trajectory, 
corresponding to the axial vector meson and lies in the limits
\begin{equation}\label{r2.2}
-0.5\leq\alpha_{A_1}(0)\leq 0
\end{equation}
This low value of the intercept (\ref{r2.2}) implies the nonsingular, flat
behaviour of the $g_1^{NS}$ function at small-$x$. The nonperturbative
contribution of the $A_1$ Regge pole is however overridden by the
perturbative QCD contributions, particularly by resummation of double
logarithmic terms $ln^2x$. In this way the Regge behaviour of the spin
dependent structure functions is unstable against the perturbative QCD
expectations, which at low-$x$ generate more singular $x$ dependence than
that implied by (\ref{r2.1})-(\ref{r2.2}). Nowadays it is well known that
the small-$x$ behaviour of the nonsinglet polarised structure function
$g_1^{NS}$ is governed by the double logarithmic terms i.e. 
$(\alpha_s ln^2x)^n$ \cite{b3},\cite{b4}. Effects of these $ln^2x$ approach go 
beyond the standard LO and even NLO $Q^2$ evolution of the spin dependent 
parton distributions and significantly modify the Regge pole model expectations 
for the structure functions. From the recent theoretical analyses of the 
low-$x$ behaviour of the $g_1^{NS}$ function \cite{b10} one can find that
resummation of the double logarithmic terms $(\alpha_s ln^2x)^n$ leads to
the singular form:
\begin{equation}\label{r2.3}
g_1^{NS}(x,Q^2) \sim x^{-\lambda}
\end{equation}
with $\lambda\approx 0.4$. This behaviour of $g_1^{NS}$ is well confirmed by
experimental data, after a low-$x$ extrapolation beyond the measured region
\cite{b2},\cite{b11},\cite{b12}.

\section{Unintegrated structure function $f^{NS}(x,Q^2)$ within double 
logarithmic $ln^2x$ and unified $ln^2x+$LO DGLAP approximations}

Perturbative QCD predicts a strong  increase of the structure function
$g_1^{NS}(x,Q^2)$ with the decreasing parameter $x$ \cite{b3},\cite{b4} what is
confirmed by experimental data \cite{b2},\cite{b11},\cite{b12}. This growth is
implied by resummation of $ln^2x$ terms in the perturbative expansion. The 
double logarithmic effects come from the ladder diagram with quark and gluon
exchanges along the chain. In this approximation the unintegrated nonsinglet
structure function $f^{NS}(x,Q^2)$ satisfies the following integral
evolution equation \cite{b3}:
\begin{equation}\label{r3.1}
f^{NS}(x,Q^2)=f_0^{NS}(x)+
\int\limits_x^1\frac{dz}{z}\int\limits_{Q_0^2}^{Q^2/z}
\frac{dk'^2}{k'^2} \bar{\alpha_s} f^{NS}(\frac{x}{z},k'^2)
\end{equation}
where
\begin{equation}\label{r3.2}
\bar{\alpha_s}=\frac{2\alpha_s}{3\pi}
\end{equation}
and $f_0^{NS}(x)$ is a nonperturbative contribution which has a form:
\begin{equation}\label{r3.3}
f_0^{NS}(x)=\bar{\alpha_s}\int\limits_x^1\frac{dz}{z} g_1^{0NS}(z)
\end{equation}
$g_1^{0NS}(x)$ is an input parametrisation
\begin{equation}\label{r3.4}
g_1^{0NS}(x)=g_1^{NS}(x,Q^2=Q_0^2)
\end{equation}
The unintegrated distribution $f^{NS}(x,Q^2)$ is related to the 
$g_1^{NS}(x,Q^2)$ via
\begin{equation}\label{r3.7}
f^{NS}(x,Q^2)=\frac{\partial g_1^{NS}(x,Q^2)}{\partial\ln Q^2}
\end{equation}
Eq. (\ref{r3.1}) generates the leading small-$x$ behaviour of $f^{NS}$ and
hence $g_1^{NS}$, but it is inaccurate in describing the total $Q^2$
evolution. For larger values of $x$, which are involved in the evolution
equation (\ref{r3.1}) via $\int\limits_x^1 dz$ one should take into account
$Q^2$ DGLAP evolution with complete splitting function $P_{qq}(z)$.
Therefore the double logarithmic approach should be completed by LO DGLAP
$Q^2$ evolution. Unified description of the polarised structure function
$f^{NS}(x,Q^2)$ incorporating DGLAP evolution and the double logarithmic
$ln^2x$ effects at low-$x$ leads to the following equation for the unintegrated 
distribution $f^{NS}(x,Q^2)$ \cite{b19}:
\begin{eqnarray}\label{r3.8}
f^{NS}(x,Q^2)&=&f_{0}^{NS}(x)+
\int\limits_x^1\frac{dz}{z}\int\limits_{Q^2}^{Q^2/z}
\frac{dk'^2}{k'^2} \bar{\alpha_s} f(\frac{x}{z},k'^2)\nonumber\\
&+&\int\limits_{Q_0^2}^{Q^2}\frac{dk'^2}{k'^2}
\int\limits_x^1\frac{dz}{z} \bar{\alpha_s}
\frac{(1+z^2)f(x/z,k'^2)-2zf(x,k'^2)}{1-z}\nonumber\\
&+&\bar{\alpha_s}\int\limits_{Q_0^2}^{Q^2}\frac{dk'^2}{k'^2}
(\frac{3}{2}+2\ln (1-x))f(x,k'^2)
\end{eqnarray}
where
\begin{eqnarray}\label{r3.9}
f_0^{NS}(x)&=&\bar{\alpha_s}[\int\limits_x^1\frac{dz}{z}
\frac{(1+z^2)g_1^{(0)}(x/z)-2zg_1^{(0)}(x)}{1-z}\nonumber\\
&+&(\frac{3}{2}+2\ln (1-x))g_1^{(0)}(x)]
\end{eqnarray}
The unintegrated distribution $f^{NS}$ in the equation (\ref{r3.8}) is related 
to the $g_1^{NS}(x,Q^2)$ via
\begin{equation}\label{r3.9a}
g_1^{NS}(x,Q^2)=g_1^{0NS}(x)+\int\limits_{Q_0^2}^{Q^2(1/x-1)}
\frac{dk^2}{k^2}f\left(x(1+\frac{k^2}{Q^2}),k^2\right)
\end{equation}
An important role in solutions of (\ref{r3.1}) and (\ref{r3.8}) plays the
coupling $\alpha_s$, which can be parametrised in different way. The simplest 
choice of $\alpha_s$ is a constance (nonrunning) coupling:
\begin{equation}\label{r3.10a}
\alpha_s = const
\end{equation}
This simplification allows the analytical analysis of the suitable evolution 
equations for truncated and full moments of the unintegrated structure function 
$f^{NS}(x,Q^2)$ within $ln^2x$ approximation \cite{b3},\cite{b9}. The 
introduction of the running coupling effects implies $\alpha_s$ in (\ref{r3.1}) 
and (\ref{r3.8}) of a form
\begin{equation}\label{r3.10b}
\alpha_s = \alpha_s(Q^2)
\end{equation}
It has been however lately proved \cite{b10}, that dealing with a very 
small-$x$ region one should use a prescription for the running coupling in 
a form $\alpha_s = \alpha_s(Q^2/z)$. This parametrisation is theoretically 
more justified than $\alpha_s =\alpha_s(Q^2)$. Namely, the substitution 
$\alpha_s =\alpha_s(Q^2)$ is valid only for hard QCD processes, when $x\sim 1$. 
However the evolution of DIS structure functions at small-$x$ needs "more 
running" $\alpha_s$:
\begin{equation}\label{r3.10c}
\alpha_s = \alpha_s(Q^2/z)
\end{equation}
Our predictions for $g_1^{NS}$ and $\Delta I_{BRS}(x_1,x_2,Q^2)$ for
different forms of $\alpha_s$ will be presented in the forthcoming section.

\section{Predictions for $g_1^{NS}$ and small-$x$ contribution to the BSR}

Our purpose is to calculate the nonsinglet polarised structure function 
$g_1^{NS}(x,Q^2)$ and hence also the contribution to the Bjorken sum rule in
the small-$x$ region. The BSR is a fundamental rule and must be hold as a 
rigorous prediction of QCD in the limit of the infinite momentum transfer 
$Q^2$:
\begin{equation}\label{r4.1}
I_{BSR} \equiv \Gamma_1^p - \Gamma_1^n =  
\int\limits_{0}^{1} dx g_1^{NS}(x,Q ^2) = \frac {1}{6}|{\frac{g_A}{g_V}}|
\end{equation}
where
\begin{equation}\label{r4.2}
\Gamma_1^p \equiv \int\limits_{0}^{1} dx g_1^{p}(x,Q ^2)
\end{equation}
\begin{equation}\label{r4.3}
\Gamma_1^n \equiv \int\limits_{0}^{1} dx g_1^{n}(x,Q ^2)
\end{equation}
and $|{\frac{g_A}{g_V}}|$ is the neutron $\beta$-decay constant
\begin{equation}\label{r4.4}
|{\frac{g_A}{g_V}}| = F + D = 1.2670
\end{equation}
Hence the BSR for the flavour symmetric sea quarks scenario
($\Delta\bar{u}=\Delta\bar{d}$) reads:
\begin{equation}\label{r4.5}
I_{BSR}(Q^2) \equiv \int\limits_{0}^{1} dx g_1^{NS}(x,Q ^2) \approx 0.211
\end{equation}
The small-$x$ contribution to the BSR has a form:
\begin{equation}\label{r4.6}
\Delta I_{BSR}(x_1,x_2,Q^2) \equiv \int\limits_{x_1}^{x_2} dx g_1^{NS}(x,Q ^2)
\end{equation}
Below we present our results for $g_1^{NS}$ and $\Delta I_{BSR}$ at small-$x$
obtained for different $\alpha_s$ sets (\ref{r3.10a})-(\ref{r3.10c}) within
combined $ln^2x+$LO DGLAP approach. We compare these predictions with pure
LO DGLAP and pure $ln^2x$ results as well. We solve numerically the
evolution equation (\ref{r3.8}) in a case of unified $ln^2x+$LO DGLAP
picture and in a case of pure LO DGLAP, when one gets the following
equation:
\begin{eqnarray}\label{r4.6a}
f^{NS}(x,Q^2)&=&f_{0}^{NS}(x)+\nonumber\\
\int\limits_{Q_0^2}^{Q^2}\frac{dk'^2}{k'^2}
\int\limits_x^1\frac{dz}{z} \bar{\alpha_s}
\frac{(1+z^2)f(x/z,k'^2)-2zf(x,k'^2)}{1-z}\nonumber\\
+\bar{\alpha_s}\int\limits_{Q_0^2}^{Q^2}\frac{dk'^2}{k'^2}
(\frac{3}{2}+2\ln (1-x))f(x,k'^2)
\end{eqnarray}
In order to have comparable results, for pure $ln^2x$ analysis we also use 
numerical solutions of (\ref{r3.1}).
Our predictions have been found for two different input parametrisations 
$g_1^{0NS}(x)$, chosen at $Q_0^2=1 {\rm GeV}^2$:
\begin{equation}\label{r4.14}
1.~~~~~g_1^{0NS}(x) = 0.8447(1-x)^3
\end{equation}
\begin{equation}\label{r4.15}
2.~~~~~g_1^{0NS}(x) = 0.290 x^{-0.4}(1-x)^{2.5}
\end{equation}
Input 1 is the simple Regge form, constance as $x\rightarrow 0$; input 2
is a "toy" model, in which we have used the latest theoretical results 
concerning the small-$x$ behaviour $x^{-0.4}$ of the nonsinglet function 
$g_1^{NS}$ \cite{b10}. In Fig.1 we plot inputs $g_1^{0NS}(x)$ (\ref{r4.14})-
(\ref{r4.15}) in the low-$x$ region $[10^{-5}\div 10^{-2}]$ together with
$g_1^{NS}(x,Q^2)$ results for $Q^2=10 {\rm GeV}^2$ within $ln^2x+$LO DGLAP
approach with "very running" coupling $\alpha_s=\alpha_s(Q^2/z)$. Fig.2
contains comparison of three approximations: pure $ln^2x$, pure LO DGLAP and
unified $ln^2x+$LO DGLAP. We present structure function $g_1^{NS}(x,Q^2=10)$
for $\alpha_s=\alpha_s(Q^2/z)$. Finally, Fig.3 shows the $g_1^{NS}(x,Q^2=10)$ 
results within combined $ln^2x+$LO DGLAP approach for different
parametrisations of $\alpha_s$: (\ref{r3.10a})-(\ref{r3.10c}). In each
figure we present the solutions for both input parametrisations
(\ref{r4.14})-(\ref{r4.15}). Numbers at each plot correspond to the suitable 
inputs 1 or 2. 
%******************Figs.1-3*************************************************
\begin{figure}[ht]
\begin{center}
\includegraphics[width=90mm]{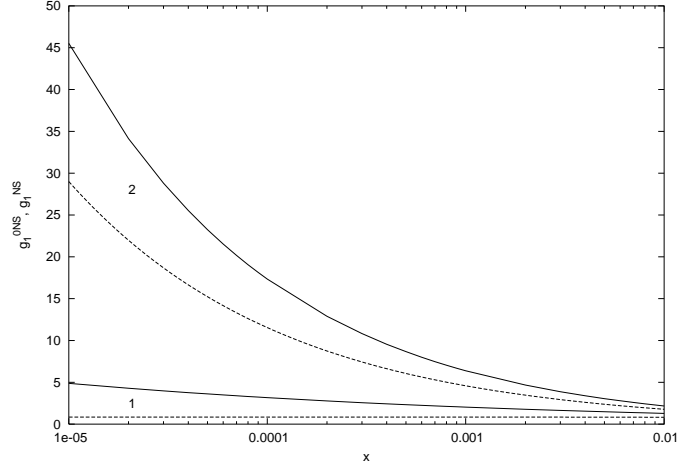}
\caption{Input parametrisations $g_1^{0NS}$ (\ref{r4.14})-(\ref{r4.15})
(dashed) and $g_1^{NS}(x,Q^2=10)$ (solid) for these inputs within 
$ln^2x+$LO DGLAP approach and for running $\alpha_s(Q^2/z)$.}
\end{center}
\end{figure}
%*****************
\begin{figure}[ht]
\begin{center}
\includegraphics[width=90mm]{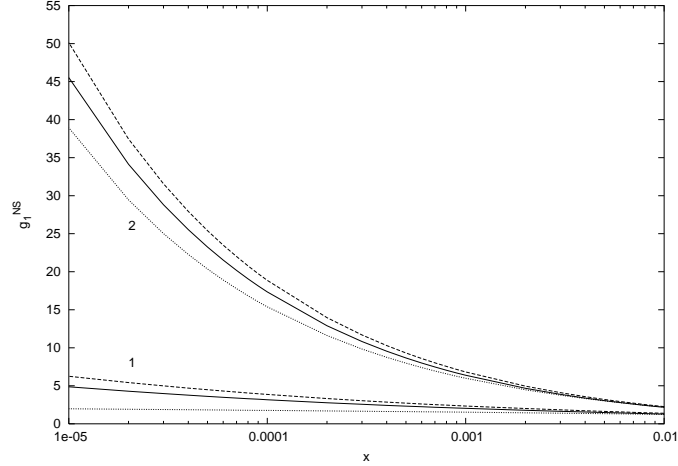}
\caption{The small-$x$ predictions for $g_1^{NS}(x,Q^2=10)$ within different
approximations: LO DGLAP (dotted), $ln^2x$ (dashed), unified $ln^2x+$LO DGLAP 
(solid). Plots for both inputs (\ref{r4.14})-(\ref{r4.15}) and running 
$\alpha_s(Q^2/z)$.}
\end{center}
\end{figure}
%*****************
\begin{figure}[ht]
\begin{center}
\includegraphics[width=90mm]{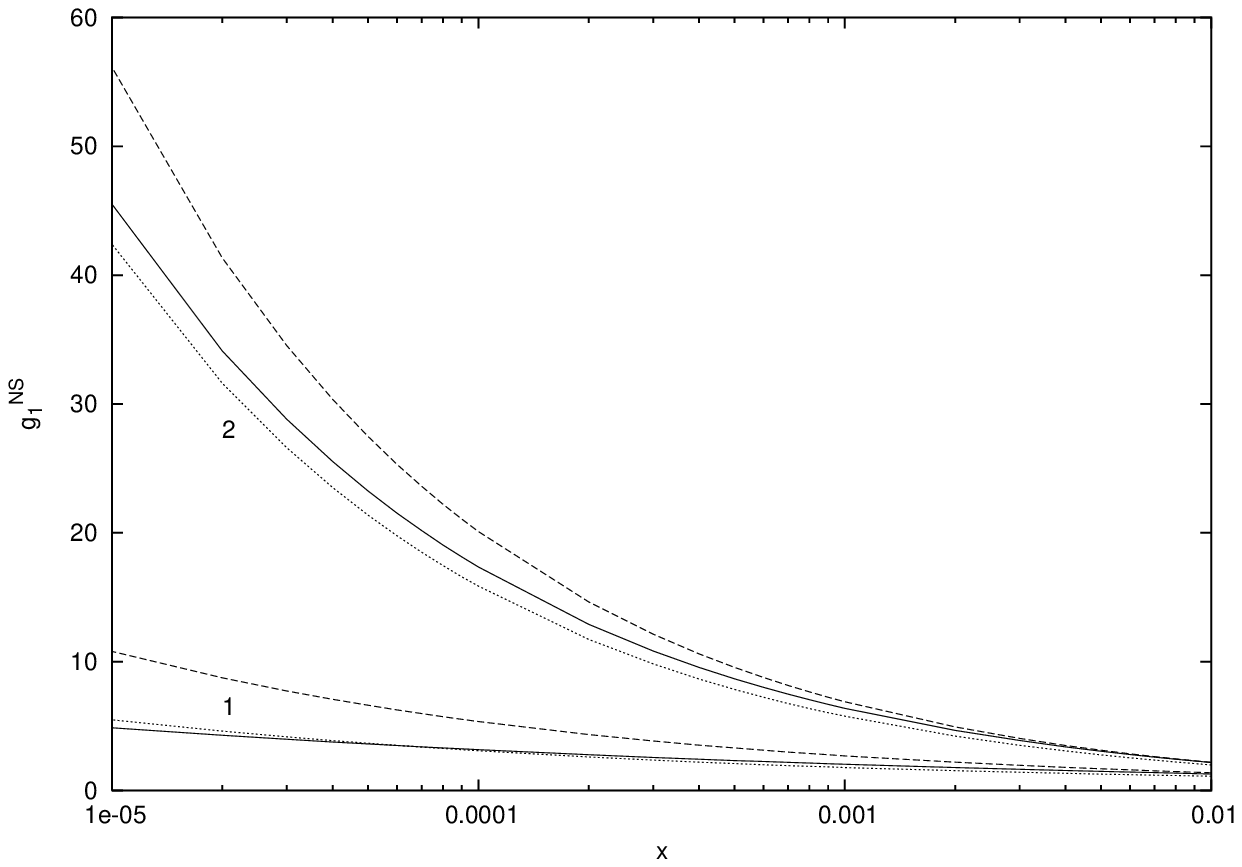}
\caption{The small-$x$ predictions for $g_1^{NS}(x,Q^2=10)$ within unified 
$ln^2x+$LO DGLAP approach for different $\alpha_s$: $\alpha_s=0.18$ (dotted),
$\alpha_s(Q^2)$ (dashed), $\alpha_s(Q^2/z)$ (solid). Plots for both inputs 
(\ref{r4.14})-(\ref{r4.15}).}
\end{center}
\end{figure}
In Table I we present our results for the low-$x$ contributions 
to the BSR (\ref{r4.6}) together with $\varepsilon (x_1,x_2)$, which is defined 
by the following expression:
\begin{equation}\label{r4.12}
\int\limits_{x_1}^{x_2} dx g_1^{NS}(x,Q ^2) = 
[1+\varepsilon (x_1,x_2)] \int\limits_{x_1}^{x_2} dx g_1^{0NS}(x)
\end{equation}
In the last column we give the percentage value $p [\%]$:
\begin{equation}\label{r4.13}
p = \frac{\Delta I_{BSR}(x_1,x_2,Q^2)}{I_{BSR}(Q^2)}\cdot 100\%
\end{equation}
In Table II we collect   results for all possible combinations of 
approximations and $\alpha_s$ sets: LO DGLAP and $\alpha_s=const=0.18$, 
LO DGLAP and $\alpha_s=\alpha_s(Q^2)$,..., $ln^2x$ and 
$\alpha_s=\alpha_s(Q^2/z)$ etc. We present here 
$\Delta I_{BSR}(0, 3\cdot 10^{-3}, 10)$, $g_1^{NS}(x=10^{-6}, Q^2=10)$. In 
the last column the effective slope $\lambda$ (\ref{r2.3}) at 
$Q^2=10 {\rm GeV}^2$ and small-$x$ $[10^{-6};10^{-5}]$ is shown. We use again 
both inputs $g_1^{0NS}(x)$.
%***Table I***
\begin{table}[ht]
\begin{center}
\begin{tabular}{|c|c|c|c|c|}
\hline\hline
$x_1$ & $x_2$ & $\Delta I_{BSR}(x_1,x_2,10)$ & $\varepsilon (x_1,x_2)$ & p\% \\
 \hline\hline
         &                  & (1) 0.006108 & 1.4213 & 2.89  \\ \cline{3-5}
   0     & $3\cdot 10^{-3}$ & (2) 0.020668 & 0.4097 & 9.80  \\ \hline
         &                  & (1) 0.016050 & 0.9289 & 7.61  \\ \cline{3-5}
   0     & $10^{-2}$        & (2) 0.040000 & 0.3336 & 18.96 \\ \hline
         &                  & (1) 0.002457 & 1.9422 & 1.16  \\ \cline{3-5}
$10^{-5}$& $10^{-3}$        & (2) 0.010450 & 0.4574 & 4.95  \\ \hline
         &                  & (1) 0.015672 & 0.9028 & 7.43  \\ \cline{3-5}
$10^{-4}$& $10^{-2}$        & (2) 0.037380 & 0.3214 & 17.72 \\ \hline
\end{tabular}
\caption{The small-$x$ contribution to the BSR (\ref{r4.6}) for different
input parametrisations (\ref{r4.14})-(\ref{r4.15}) within $ln^2x+$LO DGLAP
approximation with running $\alpha_s(Q^2/z)$.}
\end{center}
\end{table}
%***Table II***
\begin{table}[ht]
\begin{center}
\begin{tabular}{|c|c|c|c|c|c|}
\hline\hline
$g_1^{0NS}(x)$ & $approach$ & $\alpha_s$ & $\Delta I_{BSR}(0,0.003)$ &
$g_1^{NS}$ & $\lambda$ \\ \hline\hline
         &            & const=0.18        & 0.003879 & 2.07 & 0.06 \\
\cline{3-6}
         & LO         & $\alpha_s(Q^2)$   & 0.005742 & 4.31 & 0.11 \\
\cline{3-6}
         &            & $\alpha_s(Q^2/z)$ & 0.004534 & 2.17 & 0.04 \\
\cline{2-6}
         &            & const=0.18        & 0.005614 & 10.2 & 0.25 \\
\cline{3-6}
1.Regge  & $ln^2x$    & $\alpha_s(Q^2)$   & 0.008855 & 23.9 & 0.31 \\
\cline{3-6}
         &            & $\alpha_s(Q^2/z)$ & 0.007043 & 9.93 & 0.20 \\
\cline{2-6}
         &            & const=0.18        & 0.005440 & 9.75 & 0.25 \\
\cline{3-6}
         & $ln^2x$+LO & $\alpha_s(Q^2)$   & 0.008281 & 21.8 & 0.30 \\
\cline{3-6}
         &            & $\alpha_s(Q^2/z)$ & 0.006108 & 7.36 & 0.18 \\ \hline
         &            & const=0.18        & 0.017283 & 88.6 & 0.40 \\
\cline{3-6}
         & LO         & $\alpha_s(Q^2)$   & 0.020543 & 110  & 0.40 \\
\cline{3-6}
         &            & $\alpha_s(Q^2/z)$ & 0.019184 & 98.0 & 0.40 \\
\cline{2-6}
         &            & const=0.18        & 0.019026 & 113  & 0.42 \\
\cline{3-6}
2."Toy"  & $ln^2x$    & $\alpha_s(Q^2)$   & 0.023641 & 161  & 0.44 \\
\cline{3-6}
         &            & $\alpha_s(Q^2/z)$ & 0.022175 & 130  & 0.41 \\
\cline{2-6}
         &            & const=0.18        & 0.018756 & 111  & 0.42 \\
\cline{3-6}
         & $ln^2x$+LO & $\alpha_s(Q^2)$   & 0.022712 & 152  & 0.43\\
\cline{3-6}
         &            & $\alpha_s(Q^2/z)$ & 0.020668 & 117  & 0.41\\ \hline\hline 
\end{tabular}
\caption{$\Delta I_{BSR}(0,0.003,10)$, $g_1^{NS}(x=10^{-6},10)$ and $\lambda$ 
for both input parametrisations (\ref{r4.14})-(\ref{r4.15}) within different 
approaches and $\alpha_s$.}
\end{center}
\end{table}

\noindent
From these results one can read that the low-$x$ $g_1^{NS}(x,Q^2)$ values and 
hence the low-$x$ contributions to the BSR strongly depends on the input 
parametrisation $g_1^{0NS}$. For the flat Regge form (\ref{r4.14}) 
$\Delta I_{BSR}(0,10^{-2},10)$ is equal to around $7.6\%$ of the total 
$I_{BSR}=0.211$, while for the singular input (\ref{r4.15}) $19.0\%$. The
structure function $g_1^{NS}$ itself at very small-$x=10^{-6}$ and 
$Q^2=10 {\rm GeV}^2$ is in a case of $x^{-0.4}$ input about 16 times larger
than for the flat one. The value of $\varepsilon (x_1,x_2)$, defined in 
(\ref{r4.12}) varies from $0.9\div 1.9$ for the Regge input 1 to $0.3\div 0.5$ 
for the singular input 2. The effective slope $\lambda$ (\ref{r2.3})
describing the small-$x$ behaviour of the structure function $g_1^{NS}$
remains unchanged in a case of the singular input. Namely, the $x^{-0.4}$
shape of the input $g_1^{0NS}(x)$ implies again the same low-$x$ behaviour
of the $g_1^{NS}(x,Q^2)$, independently of the $Q^2$-evolution approach.
Quite different situation occurs for the flat inputs e.g. the Regge one
(\ref{r4.14}), where the singular small-$x$ behaviour of the $g_1^{NS}(x,Q^2)$
is totally generated by the QCD evolution with $ln^2x$ terms. Pure double
logarithmic $ln^2x$ approach or combined $ln^2x+$LO DGLAP approximation give
the value of $\lambda$ from 0.2 to 0.3. Only in the pure LO DGLAP analysis
we obtain $\lambda\leq 0.1$. 
It means that the double logarithmic $ln^2x$ effects are better visible in a 
case of nonsingular inputs. In a case of singular input parametrisations 
$g_1^{0NS}\sim x^{-\lambda}$ (e.g. $\lambda\sim 0.4$) the growth of 
$g_1^{NS}$ at small-$x$, implied by the $ln^2x$ terms resummation, is hidden 
behind the singular behaviour of $g_1^{0NS}$, which survives the QCD evolution. 
Comparing plots from Fig.2 and results from the last column in Table II, one
can read that the $ln^2x$ resummation gives steep growth of the $g_1^{NS}$
in the small-$x$ region. It is well visible in a case of the Regge input,
where $g_1^{NS}(x,Q^2)$ within $ln^2x$ or $ln^2x+$LO DGLAP approaches
strongly dominate over that, obtained in pure LO DGLAP approximation. Double
logarithmic contributions of the type $(\alpha_s ln^2x)^n$, which lead to
the strong growth of structure functions at low-$x$ are not included in the
DGLAP evolution (LO or NLO). Differences between pure $ln^2x$ and $ln^2x+$LO
DGLAP results within the same set of $\alpha_s$ are not very significant.
However, pure $ln^2x$ approximation overestimate the value of $g_1^{NS}$ and
should be accompanied by the LO DGLAP evolution. This is because in the
larger-$x$ region, involved in the evolution equation for $f^{NS}$ 
(\ref{r3.1}), pure $ln^2x$ analysis is inadequate. The crucial point in QCD
analysis is a treatment of the coupling $\alpha_s$. The problem of $\alpha_s$
parametrisations in high energy processes has been widely discussed in 
\cite{b10}. Fixed, constance $\alpha_s$ is very convenient in many physical
problems. Thus, use of the fixed $\alpha_s$ simplifies the evolution
equations for structure functions and enables easy analytical solutions.
However, an introduction of the fixed coupling needs a reasonable scale and
this is somehow "artificial". Namely, the scale for fixing of $\alpha_s$ is
not well defined. In perturbative QCD one should take into account running
$\alpha_s$ effects. In this way, usually, the prescription for the running
coupling reads $\alpha_s=\alpha_s(Q^2)$. This construction is however
accurate only for hard QCD processes, where $x\sim 1$. On the other hand,
many interesting QCD processes (e.g. DIS at low-$x$) are Regge-like. For
these cases, with small-$x$ involved, "hard" running $\alpha_s(Q^2)$ is
incorrect. Instead of $\alpha_s(Q^2)$ one has to use a modified 
parametrisation of $\alpha_s$:
\begin{equation}\label{r4.17}
\alpha_s = \alpha_s(k_{\perp}^2/\beta)
\end{equation}
where $k_{\perp}^2$ is the transverse momentum of the ladder parton and $\beta$ 
is the standard Sudakov parameter. In our approach this prescription reads
as (\ref{r3.10c}). From Fig.3 and Table II we are able to compare the
predictions for $g_1^{NS}$ at small-$x$ for three $\alpha_s$ parametrisations 
(\ref{r3.10a})-(\ref{r3.10c}). The results for $\alpha_s=const=0.18$ and 
$\alpha_s=\alpha_s(Q^2/z)$ (within the same approach LO or $ln^2x$ etc. and 
with the same input $g_1^{0NS}$) are similar but significantly smaller than
in a case of running $\alpha_s=\alpha_s(Q^2)$. Within $ln^2x+$LO DGLAP
approximation with flat input for "very" running $\alpha_s(Q^2/z)$ ($x<z<1$), 
via weaker coupling ($\alpha_s(Q^2/z)<\alpha_s(Q^2)$), the value of 
$g_1^{NS}(x=10^{-6},Q^2=10)$ is almost 3 times smaller than for the "hard"
running $\alpha_s(Q^2)$. It is a good lesson how choice of the running
coupling influences the results in the low-$x$ region. 
From the experimental SMC data \cite{b11} the low-$x$ contribution to the BSR 
at $Q^2=10 {\rm GeV}^2$ is equal to
\begin{equation}\label{r4.18}
6 \int\limits_{0}^{0.003} g_1^{NS}(x,Q ^2=10) dx = 0.09\pm 0.09
\end{equation}
The above result has been obtained via an extrapolation of $g_1^{NS}$ to the
unmeasured region of $x$: $x\rightarrow 0$. Forms of the polarised quark
distributions have been fitted to SMC semi-inclusive and inclusive
asymmetries. In the fitting different parametrisations of the polarised
quark distributions \cite{b16} \cite{b17} have been used. The extrapolation of 
$g_1^{NS}$ to very small-$x$ region depends strongly on the assumption (input 
parametrisation) made for this extrapolation. In this way
present experimental data give only indirectly the estimation of the
small-$x$ contribution to the moments of parton distributions. The result 
(\ref{r4.18}) with a large statistical error and strongly fit-dependent
cannot be a final, crucial value. Nevertheless we would like to estimate the
exponent $\lambda$ in the low-$x$ behaviour of $g_1^{NS}\sim x^{-\lambda}$
using the above SMC result for the small-$x$ contribution to the BSR.
Assuming the validity of the BSR (\ref{r4.5}) at large $Q^2=10 {\rm GeV}^2$,
one can find:
\begin{equation}\label{r4.19}
\int\limits_{0}^{x_0} dx g_1^{NS}(x,Q ^2) = 
I_{BSR}(Q^2) - \int\limits_{x_0}^{1} dx g_1^{NS}(x,Q ^2)
\end{equation}
where $x_0$ is a very small value of the Bjorken variable. Taking into
account the small-$x$ dependence of $g_1^{NS}\sim x^{-\lambda}$ and the
experimental data for $\Delta I_{BSR}(0,0.003,10)$ one can obtain:
\begin{equation}\label{r4.20}
C \int\limits_{0}^{0.003} x^{-\lambda} dx = 0.015\pm 0.015
\end{equation}
The constant $C$ can be eliminated from a low-$x$ SMC data \cite{b11}:
\begin{equation}\label{r4.21}
C x^{-\lambda} = g_1^{n-p}(x,10)
\end{equation}
Taking different small-$x$ SMC data, we have found $\lambda =0.37$
($x=0.014$); $\lambda =0.20$ ($x=0.008$); $\lambda =0.38$ ($x=0.005$). 

It seems nowadays that the most probably small-$x$ behaviour of $g_1^{NS}$ is
\begin{equation}\label{r4.22}
g_1^{NS}(x,Q^2) \sim x^{-0.4}
\end{equation}
This results from latest theoretical analyses \cite{b10}, which take into
account the running coupling effects at low-$x$. It has been shown
in \cite{b10} that the intercept $\lambda$ controlling the power-like small-$x$ 
behaviour of $g_1^{NS}$ depends on the choice of the parameters $n_f$ (flavour 
number), $Q_0^2$ (input scale) and $\Lambda_{QCD}$. The maximal value of
$\lambda$, which gives the maximal contribution to the structure function
$g_1^{NS}$ in the perturbative QCD, incorporating $ln^2x$ effects, is equal to 
0.4. The same value of $\lambda =0.4$ was obtained in the semi-phenomenological 
estimation from BSR for lower $Q^2$ \cite{b18}. 
The polarised (nonsinglet and singlet as well) structure functions are
presently the objects of intensive theoretical investigations. Maybe the 
crucial point for understanding of the small-$x$ behaviour of structure 
functions is an analysis beyond the leading order $\alpha_s^n ln^{2n}x$. 
The resummation of $\alpha_s^{(n+1)} ln^{2n}x$ terms is studied in \cite{b21}.
The corrections to $g_1^{NS}$ due to the nonleading terms are on the level
of 1$\%$ in the accessible at present experimentally $x$ region, but can be
larger (up to about 15$\%$) at very small-$x\sim 10^{-5}$. In the situation, 
when the present experimental data do not cover the whole region of $x\in (0;1)$, 
theoretical predictions for e.g. structure functions in the unmeasured low-$x$ 
region cannot be directly verified. Latest experimental SMC \cite{b2},
\cite{b11} and HERMES \cite{b12} data provide results for the BSR from the 
region $0.003\leq x\leq 0.7$ and $0.023\leq x\leq 0.6$ respectively. In the 
very small-$x$ region exist only indirect, extrapolated results with large 
uncertainties. Small-$x$ contribution to the Bjorken sum rule resulting from 
such indirect SMC data analysis is equal to $0.015 \pm 0.015$. Large 
uncertainties of the small-$x$ experimental results disable unfortunately 
realistic comparison the data with the theoretical predictions. Namely, all 
our results for $\Delta I_{BSR}(0,0.003,10)$ in Table II: from 0.004 
(for LO, $\alpha_s=0.18$, flat input) to 0.024 (for $ln^2x$, $\alpha_s(Q^2)$, 
singular $x^{-0.4}$ input) are in agreement with SMC data (within the total 
error). Nevertheless, the progress in theoretical \cite{b3},\cite{b4},
\cite{b10},\cite{b19},\cite{b20},\cite{b21} and experimental \cite{b2},
\cite{b11},\cite{b12} investigations gives hope that our knowledge about 
structure functions at small-$x$ is getting better.

\section{Summary and conclusions}

In this paper we have estimated the nonsinglet polarized structure function
$g_1^{NS}$ at small-$x$ and also contributions from the small-$x$ region
to the Bjorken sum rule. We have used the numerical solutions within unified 
double logarithmic  and DGLAP ($ln^2x+$LO DGLAP) approximation. Our
predictions for $g_1^{NS}(x,Q^2)$ and $\Delta I_{BSR}(x_1,x_2, Q^2)$ have 
been found for two input parametrisations $g_1^{0NS}(x,Q_0^2)$.
These parametrisations describe different small-$x$ behaviour
of $g_1^{0NS} = g_1^{0(p-n)}$ at $Q_0^2$: $g_1^{NS}\sim x^{-\lambda}$. 
The main conclusion from our analyses is that the structure function 
$g_1^{NS}$ at small-$x$ and hence also the small-$x$ contribution to the BSR 
strongly depends on the input parametrisation $g_1^{0NS}$. 
The percentage value $\Delta I_{BSR}(0,10^{-2},Q^2 = 10)$ of the total BSR 
$\approx 0.211$ varies from 7.6 for the flat Regge input 1 ($\lambda =0$) to 
almost 19 for the singular one 2 ($\lambda =0.4$). The structure function 
$g_1^{NS}$ itself at very small-$x=10^{-6}$ and $Q^2=10 {\rm GeV}^2$ is in 
a case of $x^{-0.4}$ input about 16 times larger than for the flat one. 
Double logarithmic $ln^2x$ effects, responsible for the strong growth of the
structure function in the low-$x$ region, are better visible in a 
case of nonsingular inputs. In a case of singular input parametrisations 
$g_1^{0NS}\sim x^{-\lambda}$ (e.g. $\lambda\sim 0.4$) the growth of 
$g_1^{NS}$ at small-$x$, implied by the $ln^2x$ terms resummation, is hidden 
behind the singular behaviour of $g_1^{0NS}$, which survives the QCD evolution. 
Input parametrisation 2 incorporates latest theoretical investigations,
which suggest singular small-$x$ shape of polarised structure functions: 
$\sim x^{-0.4}$ for the nonsinglet case and even $\sim x^{-0.8}$ for the 
singlet one. Both these values are indirectly confirmed by fitted experimental 
HERMES data. Basing on these results, similar extrapolations of the spin 
dependent quark distributions towards the very low-$x$  region have been 
assumed in several recent input parametrisations $\Delta q(x,Q_0^2)$.
Our results for the small-$x$ contribution $0\leq x\leq 0.003$ to the BSR
are in agreement with the experimental SMC data (for both inputs). However it 
must be emphasized, that SMC data for the low-$x$ region suffer from large 
uncertainties. Using SMC data for $g_1^{NS}$ at small-$x$ ($0.14$, 
$5\cdot 10^{-3}$, $8\cdot 10^{-3}$) we have estimated the exponent $\lambda$ 
which governs the low-$x$ behaviour of $g_1^{NS}$. Thus we have obtained 
$\lambda =0.20\div 0.38$ with large uncertainties. This effective slope 
$\lambda$ calculated for low-$x\in [10^{-6};10^{-5}]$ from $g_1^{NS}$ in our 
approach amounts about 0.2 (for running $\alpha_s(Q^2/z)$ and Regge-like flat 
input).
In order to have reliable theoretical predictions for the polarised structure 
function $g_1^{NS}(x,Q^2)$ we have used unified approach which 
contains the resummation of the $ln^2x$ and the LO DGLAP $Q^2$ evolution as 
well. It is because the pure $ln^2x$ approximation generates correctly the 
leading small-$x$ behaviour of the polarised structure function but is 
inaccurate for larger values of $x$. Another crucial point of the presented
analysis is the role of the running coupling effects. 
Latest theoretical studies suggest introduction of the running coupling of 
a form $\alpha_s=\alpha_s(Q^2/z)$ instead of $\alpha_s=\alpha_s(Q^2)$. 
This is more justified in the small-$x$ region. We have found that the choice 
of the running coupling significantly influences the results in the low-$x$ 
region. E.g. the value of $g_1^{NS}(x=10^{-6},Q^2=10)$ is for "hard" running 
$\alpha_s(Q^2)$ almost 3 times greater than for the "very" running 
$\alpha_s(Q^2/z)$ (in a case of nonsingular input $g_1^{0NS}$).
Proper theoretical treatment of the $Q^2$ evolution of structure functions
in the whole (small and large) $x$ region with all essential perturbative
leading and even nonleading effects involved should be a subject of further 
intensive investigations. It is important because of lack of the experimental 
data from the very small-$x$ region ($x<0.003$). Agreement of the theoretical 
predictions e.g. for the BSR with real experimental data at medium and large 
$x$ may give hope, that for the very interesting small-$x$ region the 
suitable theoretical results are also reliable.

\section*{Acknowledgements}

We thank Boris Ermolaev for constructive remarks and useful comments
concerning the running coupling effects in the small-$x$ region. We are also
grateful to Johannes Bl\"umlein for pointing out the role of nonleading terms 
in polarised structure functions.

\end{document}